# SESNet: sequence-structure feature-integrated deep learning method for data-efficient protein engineering

Mingchen Li[1,4†], Liqi Kang[1,2†], Yi Xiong[5], Yu Guang Wang[1], Guisheng Fan[4], Pan Tan[1*], Liang Hong[1,2,3*]

1. Shanghai National Center for Applied Mathematics (SJTU Center), & Institute of Natural Sciences, Shanghai Jiao Tong University, Shanghai 200240, China
2. School of Physics and Astronomy & School of Pharmacy, Shanghai Jiao Tong University,200240, China
3. Shanghai Artificial Intelligence Laboratory, Shanghai 200240, China
4. School of Information Science and Engineering, East China University of Science and Technology, Shanghai 200240, China
5. School of Life Sciences and Biotechnology, Shanghai Jiao Tong University, Shanghai 200240, China

## Abstract

Deep learning has been widely used for protein engineering. However, it is limited by the lack of sufficient experimental data to train an accurate model for predicting the functional fitness of high-order mutants. Here, we develop SESNet, a supervised deep-learning model to predict the fitness for protein mutants by leveraging both sequence and structure information, and exploiting attention mechanism. Our model integrates local evolutionary context from homologous sequences, the global evolutionary context encoding rich semantic from the universal protein sequence space and the structure information accounting for the microenvironment around each residue in a protein. We show that SESNet outperforms state-of-the-art models for predicting the sequence-function relationship on 26 deep mutational scanning datasets. More importantly, we propose a data augmentation strategy by leveraging the data from unsupervised models to pre-train our model. After that, our model can achieve strikingly high accuracy in prediction of the fitness of protein mutants, especially for the higher order variants (> 4 mutation sites), when finetuned by using only a small number of experimental mutation data (<50). The strategy proposed is of great practical value as the required experimental effort, i.e., producing a few tens of experimental mutation data on a given protein, is generally affordable by an ordinary biochemical group and can be applied on almost any protein.

## Introduction

Proteins are workhorses of the life activities. Their various functions such as catalysis, binding, and transportation undertake most of the metabolic activities in cells. In addition, they are the key components of the cytoskeleton, supporting the stable and diverse form of organisms. Nature provides numerous proteins with great potential value for practical applications. However, the natural proteins often do not have the optimal function to meet the demand of bioengineering. Directed evolution is a widely used experimental method to optimize proteins' functionality, namely fitness, by employing a greedy local search to optimize protein fitness[1, 2]. During this process, gain-of-function mutants are achieved and optimized via mutating several Amino Acids (AA) in the protein, which were selected and accumulated through the iterative processes of mutation by testing hundreds to thousands of variants in each generation. Despite the great success directed evolution has achieved, the phase space of the protein fitness landscape can be screened by this method is rather limited. Furthermore, to acquire a mutant of excellent fitness, especially a high-order mutant with multiple AA being mutated, the directed evolution often needs to develop an effective high-throughput screening or conduct a large number of experimental tests, which is experimentally and economically challenging[3].

Since experimental screening for directed evolution is largely costing, particularly for high-order mutations, prediction of the fitness of protein variants in silico are highly desirable. Recently, deep learning methods have been applied for predicting the fitness landscape of the protein variants[2]. By building models trained to learn the sequence-function relationship, deep learning can predict the fitness of each mutant in the whole sequence space and give a list of the most favorable candidate mutants for experimental tests. Generally, these deep learning models can be classified into protein language models [4-11], learning the representations from the global unlabeled sequences[6, 7, 12] and multiple sequence alignment (MSA) based model, capturing the feature of evolutional information within the family of the protein targeted[13-16]. And more recent works have proposed to combine these two strategies: learning on evolutionary information together with global natural sequences as the representation[17, 18], and trained the model on the labelled experimental data of screened variants to predict the fitness of all possible sequences. Nevertheless, all these models are focused on protein sequence, i.e., using protein sequence as the input of the model. Apart from sequence information, protein structure can provide additional information on function. Due to the experimental challenge of determining the protein structure, the number of reported protein structures is orders of magnitude smaller than that of known protein sequences, which hinders the development of geometric deep learning model to leverage protein structural feature. Thanks to the dramatic breakthrough in deep learning-based technique for predicting protein structure[19, 20], especially AlphaFold 2, it is now possible to efficiently predict protein structures from sequences at a large scale [21]. Recently, some researches directly take the protein structure feature as input to train the geometric deep learning model, which has been proved to achieve better or similar performance in prediction of protein function compared to language models [22-24]. However, the fused deep-learning method which can make the use of both sequence and structural information of the protein to map the sequence-function is yet much to be explored [25].

Recently, both supervised and unsupervised models have been developed for protein engineering, i.e., prediction of the fitness of protein mutants[24, 26]. Generally speaking, the supervised model can often achieve better performance as compared to the unsupervised model[26], but the former requires a great amount (at least hundreds to thousands) of experimental mutation data of the protein studied for training, which is experimentally challenging[18]. In contrast, the unsupervised model does not need any of such experimental data, but its performance is relatively worse, especially for the high-order mutant, which is often the final product of a direct-evolution project. It is thus highly desirable to develop a deep-learning algorithm, which can efficiently and accurately predict the fitness of protein variants, especially the high-order mutant, without the need of a large size of experimental mutation data of the protein concerned. In the present work, we built a supervised deep learning model (SESNet), which can effectively fuse the protein sequence and structure information together to predict the fitness of variant sequences (Fig 1A). We demonstrated that SESNet outperforms several state-of-the-art models on 26 metagenesis datasets. Moreover, to reduce the dependence of the model on the quantity of experimental mutation data, we proposed a data-augmentation strategy (Fig 1B), where the model was firstly pre-trained using a large quantity of the low-quality results derived from the unsupervised model and then finetuned by a small amount of the high-quality experimental results. We showed that the proposed model can achieve very high accuracy in predicting the fitness of high-order variants of a protein, even for those with more than four mutation sites, when the experimental dataset used for finetuning is as small as 40. Moreover, our model can predict the key AA sites, which are crucial for the protein fitness, and thus the protein engineer can focus on these key sites for mutagenesis. This can greatly reduce the experiment cost of trial and error.

## Results

**Deep learning-based architecture of SESNet for predicting protein fitness.**

To exploit the diverse information from protein sequence, coevolution and structure, we fuse three encoder modules into our model. As shown in Fig 1A: the first one (local encoder) accounts for residue interdependence in a specific protein learned from evolution-related sequences[15, 16]; the second one (global encoder) captures the sequence feature in global protein sequence universe[6, 12]; and the third one (structure module) captures local structural feature around each residue learned from 3D geometric structure of the protein[23, 24]. To integrate the information of different modules, we first concatenate representations of local and global encoders and get an integrated sequence representation. This integrated sequence representation is then sent to an attention layer and becomes the sequence attention weights, which will be further averaged with the structure attention weights derived from structure module, leading to the combined attention weights. Finally, the product of combined attention weights and the integrated sequence representation is then fed into a fully connected layer to generate the predicted fitness. The combined attention weights can also be used to predict the key AA sites, critical for the protein fitness, details of which is discussed in the section of Method.

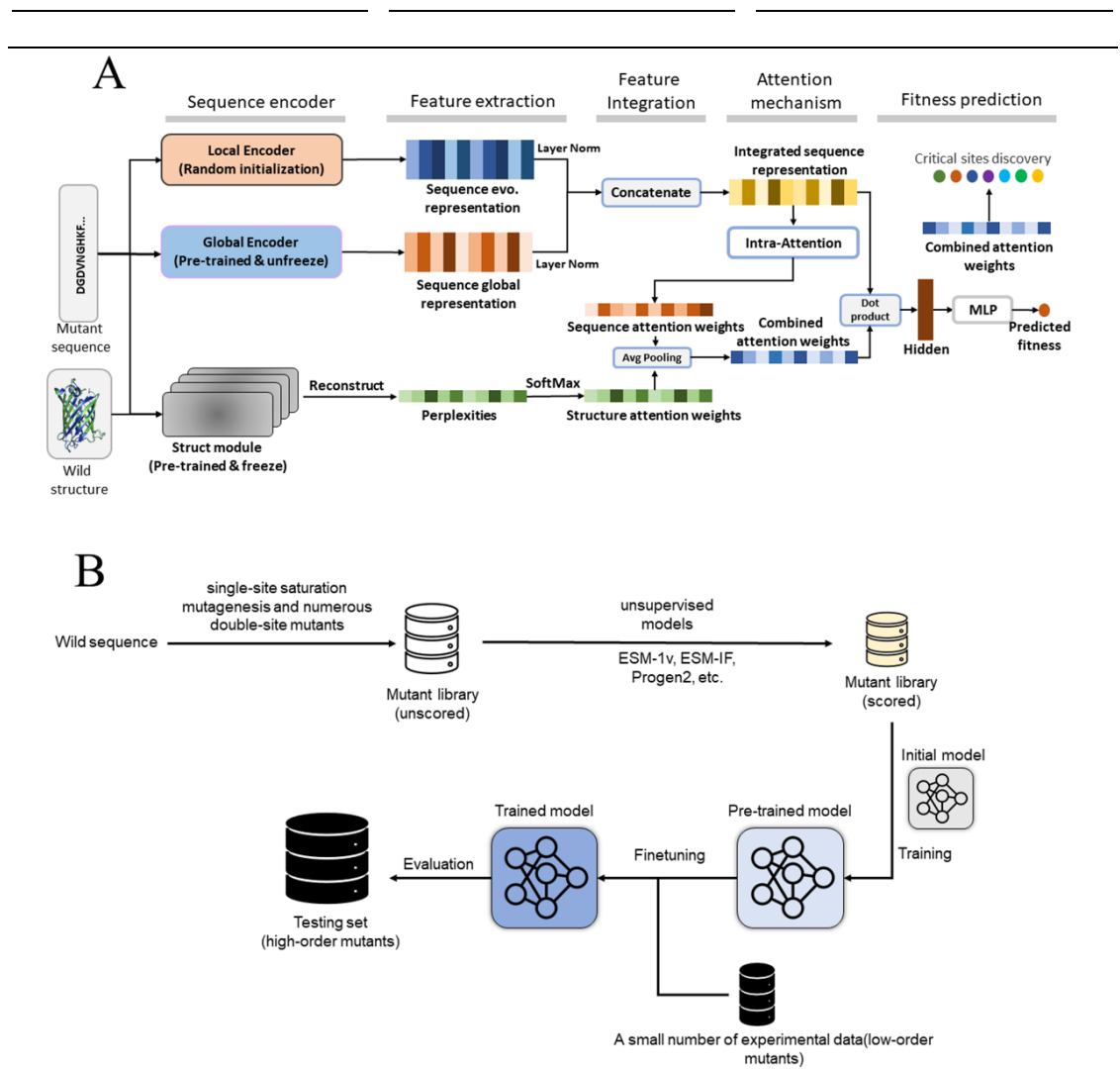

**Figure 1. Architecture of model and the schematic of data-augmentation strategy.** Architecture of SESNet (A): The local encoder accounts for the inter-residue dependence in a protein learned from MSA of homologous sequences using a Markov random field[27]. The global encoder captures the sequence feature in global protein sequence universe using protein language model[6]. The structure module accounts for the microscopically environmental feature of a residue learned from 3D geometric structure of the protein[23, 28]. Schematic of data-augmentation strategy. (B): We first build a mutant library containing all of the single-site mutants and numerous double-site mutants. Then, all of these mutated sequences are scored by the unsupervised model. After that, these mutants are used to pre-train the initial model (SESNet), which will be further finetuned on a small number of low-order experimental mutational data.

## SESNet outperforms state-of-the-art methods for predicting fitness of variants on deep mutation scan (DMS) datasets

We compared our supervised model against the existing state-of-the-art supervised models, ECNet [17], ESM-1b [6]; and unsupervised models, ESM-1v [9], ESM-IF1[23] and MSA transformer[15]. As can be seen in Fig 2A, in 19 out of 20 datasets, the supervised models generally outperform the unsupervised ones as expected, and our model (SESNet) achieves the best performance among all the models. Moreover, we further explored the ability of our model to predict the fitness of higher-order variants by training it using the experimental results of the low-order variants on 6 datasets of DMS. As shown in Fig 2B&C, our model outperforms all the other models. Data in Fig 2 is presented in Supplementary Tables 1,2&3. These datasets cover various proteins and different types of functionalities, including catalytic rate, stability, and binding affinity to peptide, DNA, RNA and antibody, as well as fluorescence intensity (Table 4). While most of the datasets contain only single-site mutants, five of them involve both single-site and double-site mutants, and the dataset of GFP contains data up to 15-site mutants.

**All three components contribute positively to the performance of SESNet.**

As described in the above architecture (Fig.1A), our model integrates three different encoders or modules together. To investigate how much contribution each of the three parts makes, we performed ablation studies in 20 datasets of single-site mutants. Briefly, we removed each of the three components and compared the performance to that of the original model. As shown in Supplementary Table 5, the average spearman correlation of the original model is 0.672, much higher than that without local encoder (0.639), that without global encoder (0.247) and that without structure module (0.630). The ablation study reveals that all three components contribute to the improvement of model performance, and the contribution from the global encoder, which captures the sequence feature in global protein sequence universe, is the most significant.

**The combined attention weights guide the finding of the key AA site.**

The combined attention weights can be used to measure the importance of each AA site on protein fitness when mutated. To the first approximation, higher the attention score is, more important the AA site is. To test this approximation, we trained our model on the experimental data of 1084 single-site mutants in the dataset of GFP [29], a green fluorescent protein from *Aequorea victoria*. The ground truth of the key sites of GFP are defined here as the experimentally discovered top 20 sites, which exhibit the largest change of protein fitness when mutated, or the AAs forming and stabilizing the chromophore, which are known to significantly affect the fluorescent function of the protein [30], but lack the fitness results in the experimental dataset. Indeed, one can observe that, at least 4 out of 7 top attention-score AA sites predicted by our model are the key sites as two of them (AG65 and T201) are located at the chromophore, and the other two (P73 and R71) were among the top 20 residues discovered in experiment to render the highest change of fitness when mutated (Fig 3A and Fig S1A). Interestingly, when we removed the structure module from the model, only one residue in the predicted top-7 attention-score AA is the key site (Fig 3B and Fig S1B).

To further verify this discovery, we also performed these tests on the dataset of RRM, the RNA recognition motif of the *Saccharomyces cerevisiae* poly(A)-binding protein[31]. The key sites of RRM are defined as the experimentally discovered top 20 sites, which render the largest change of fitness of the protein when mutated, or the binding sites, which are within 5 Å of the RNA molecules as revealed in the structure of PDB 6R5K. Fig 3C and Fig S2A show that 4 out of 7 top attention-score AA sites predicted by our model are the key AAs. One of them (I12) is among the top 20 residues and three of them (N7, P10 and K39) are binding sites. Whereas, no key residue can be found in the predicted top-seven attention-score AAs, when we removed the structure module. (Fig 3D and Fig S2B).

The results in Fig. 3 demonstrate that the structural module which learns the microscopically structural information around each residue makes important contribution to identify the key AAs, which are crucial for the protein fitness. Although the ablation study (Supplementary Table 5) reveals that the addition of the structural module improves the average spearman correlation over 20 datasets only by 4 percent, Fig. 3 demonstrates an important role of the structural module, which can guide the protein engineer to identify the important AA sites in a protein for mutagenesis .

**Data-augmentation strategy boosts the performance of the fitness prediction when finetuned by a small size of labelled experimental data.**

Supervised model is normally performing better than the unsupervised models (see Fig. 2)[26]. But the accuracy of the supervised model is highly affected by the amount of input experimental results used for training. However, it is experimentally challenging and costly to generate sufficient data (many hundreds or even thousands) for such purpose on every protein studied. To address this challenge, we propose a simple strategy of data augmentation by using the result generated by one unsupervised model to pre-train our model on a given protein, and then finetuning it using a limited number of experimental results on the same protein. We call it a pre-trained model. We note that data-augmentation strategy has been applied in various earlier work and has achieved good success in protein design[23, 32, 33]. In particular, to improve the accuracy of inverse folding, ref [23] used 16153 experimentally determined 3-D structures of proteins and 12 million structures predicted by the AlphaFold 2 [19] to train the model ESM-IF1[23] . In the present work, the data augmentation strategy is used for a different purpose that it can reduce the dependence of the supervised model on the size of the experimental data when predicting the fitness of protein mutants. We took GFP as an example to illustrate our data-augmentation strategy as GFP has a large number of experimental data for testing, particularly the experimental data for high-order mutants (up to 15-site mutant). We used the fitness results of low-order mutants predicted by the unsupervised model, ESM-IF1, to pre-train our model. The pre-training dataset contains the fitness of all single-site mutants and 30,000 double-site mutants randomly selected out of tens of million double-site variants. Then, we finetuned the pre-trained model by a certain number of experimental results of single-site mutants. The resulting model was used to predict the fitness of high-order mutants. As can be seen in Fig. 4A-D, when comparing with the original model without pre-training (blue bars), the performance of the pre-trained model is significantly improved (red bars). Such improvement is particularly large when only a small number of experimental data (40) is fed for training, and it will be gradually reduced when feeding more experimental data, eventually disappearing when more than 1000 experimental data were used for training. Here, we would like to particularly highlight the case when the finetuning experimental dataset contains only 40 data points. As can be seen in Fig. 4A, the pretrained model can achieve high spearman correlation of 0.5-0.7 for multisite-mutants, even for high-order mutants with 5-8 mutation sites. This is remarkably important for most protein engineers, as such experimental workload (40 data points) is generally affordable in an ordinary biochemical research group. However, without pre-training, the performance of the supervised model is rather low (~0.2). This comparison demonstrates the advantage of the data augmentation strategy proposed in the present work.

Moreover, we also compared the performance of the pretrained model with respect to the unsupervised model (green bars), which were used for generating the low-quality pretraining datasets. As can be seen, when only 40 experimental data were used for training, the pretrained model has similar performance as compared to the unsupervised model for low-order mutants (< 4 mutation sites), but clearly outperforms the latter for high-order mutants (>4 mutation sites). When feeding more experimental data, especially a couple of hundreds, the pretrained model will outperform the unsupervised model regardless of how many sites of the protein were mutated.

The unsupervised model used for analysis in Fig. 4 is ESM-1F1, which captures the local structural information of a residue. To demonstrate the general superiority of data-augmentation strategy proposed here, we also tested the results using other unsupervised model to generate the augmented datasets for GFP. As can be seen in Fig. S3, we used ProGen2 [8], an unsupervised model to learn the global sequence information, for data augmentation, and still derived the similar conclusion as in Fig. 4. That is, the pretrained model outperforms the original model without pretraining especially when a small experimental dataset is used for training, and it also beats the unsupervised model particularly for the high-order mutants.

To further validate the generality of the data augmentation strategy proposed here, we did the analysis on the dataset of other proteins: toxin-antitoxin complex (F7YBW8) [34]containing data up to 4 sites mutants, and Adeno-associated virus capsids (CAPSD_AAV2S)[35], a deep mutational dataset including data up to 23-site mutants. We used the unsupervised model ProGen2[8] to generate the low-quality data of F7YBW8 for pretraining, since we found ProGen2 performs better than ESM-IF1 on this dataset. As shown in Fig 5A, the pre-trained model outperforms both the original model without pretraining and the unsupervised model in the fitness prediction of all multi-site mutants (2-4 sites) after finetuned by using only 37 experimental data points. In addition, in the dataset of CAPSD_AAV2S (Fig 5B), the pre-trained model also achieves the best performance in all of the high-order mutants ranging from 2 to 23 sites, when finetuned by only 20 experimental data points. These results further support the practical use of our data augmentation strategy, as the required experimental effort is largely affordable on most proteins.

**Learned models provide insight into protein fitness.**

SESNet projects a protein sequence into a high dimensional latent space and represents each mutant as a vector by the last hidden layer. Thus, we can visualize the relationships between sequences in these latent spaces to reveal how the networks learn and comprehend protein fitness. Specifically, we trained SESNet on the experimental data of single-site mutants from the datasets of GFP and RRM, then we used the trained model and untrained model to encode each variant and extracted the output of the last hidden layer as a representation of the variant sequence. Fig S4 shows a two-dimensional projection of the high dimensional latent space using t-SNE[36]. We found that the representations of positive and negative variants, i.e., the experimental fitness values being larger or smaller than that of wildtype, generated by the trained SESNet are clearly clustered into distinct groups (Fig S4A and Fig S4B). In contrast, the representations from untrained model cannot provide a distinguishable boundary between positive and negative variants (Fig S4C and Fig S4D). Therefore, SESNet can learn to distinguish functional fitness of mutants into a latent representation space with supervised training.

Furthermore, to explore why the data-augmentation strategy works, we performed a case study on GFP dataset. Here, we compared the latent-space representation from the last hidden layer generated by our model with and without pre-training using the augmented data from the unsupervised model. As seen in Fig. S5A, after pretraining even without finetuning by the experimental data, SESNet can already roughly distinguish the negative and positive mutants. One thus can deduce that the pre-training can furnish a good parameter initialization for SESNet. After further finetuning the pre-trained SESNet by only 40 experimental data points of single-site mutants, a rather clear boundary between negative and positive high-order mutants is further outlined (Fig S5B). In contrast, when we skipped the pretraining process, i.e., directly training the model on 40 experimental data points, the separation between the positive and negative high-order mutants is rather ambiguous (Fig S5C). This comparison demonstrates the superiority of our data-augmentation strategy in distinguishing mutants of distinct fitness values, when the number of available experimental data is limited.

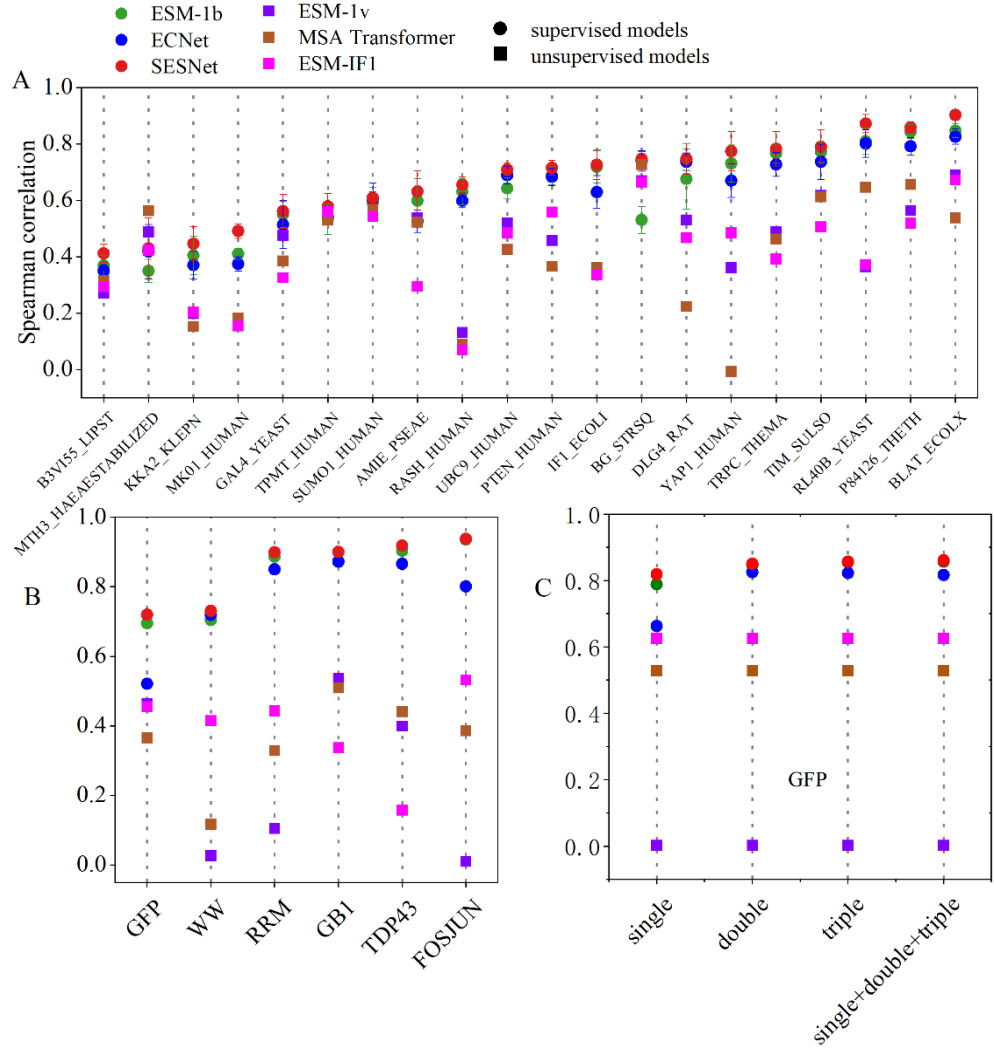

**Figure 2. Spearman correlation of predicted fitness.** A: Comparison of our model to other models on the predicted fitness of the single-site mutants on 20 datasets. We performed five-fold cross-validation with 7:1:2 as the ratio of train versus validation versus test set. B: comparison of predicted fitness of double-site mutants of our model to other unsupervised models (ESM-1v, ESM-IF1 and MSA transformer), or supervised models (ECNet and ESM-1b). Here, our model and other supervised models were trained on the data of single-site mutants. We used 10% of double-site mutants as validation set and the remaining 90% as test set. C: Comparison of our model to other models on fitness prediction of quadruple-site mutants of GFP. Here, our model and other supervised model were trained using the single, double, triple-site mutants and all the three together. We used 10% of quadruple-site mutants as validation set and the remaining 90% as test set. The error bar in single-site mutant was got from the five-fold cross-validation. Since we cannot do five-fold cross-validation in the fitness prediction of high-order mutants trained on low-order mutants, we don't put error bar for those data.

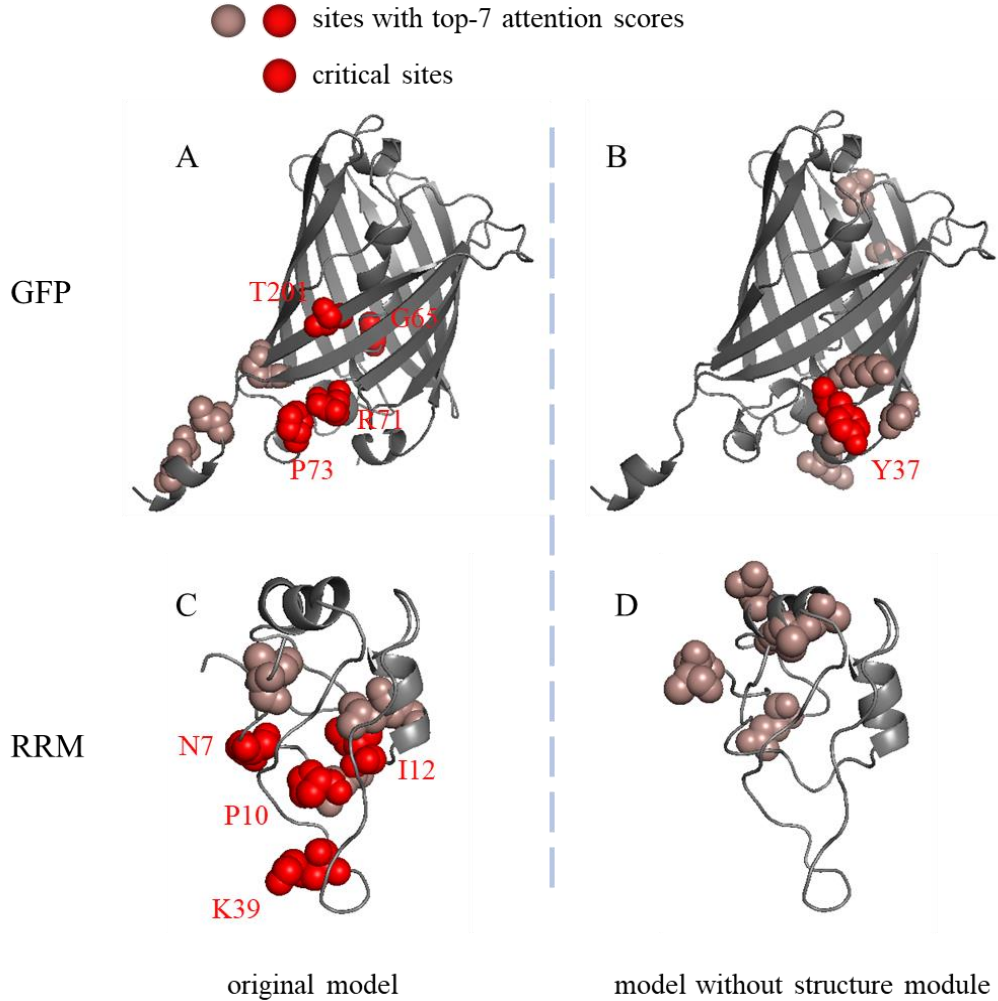

**Figure 3. The sites with the top 7 largest attention scores on the wildtype sequence.** A&B: The key sites of GFP have been marked as red spheres. A: 4 key sites were recovered by our model. G65 and T201 are the active residues helping to form and stabilize the chromophore in GFP as described by Ref [30]. P73 and R71 are among the experimentally-discovered top 20 sites, which render the highest change of fitness when mutated. B: Only one key site was identified by the model when removing the structure module and it is Y37, which is among the experimentally-discovered top 20 AA sites. C&D: The key sites of RRM have been marked as red spheres. C: 4 key sites were recovered by the original model. N7, P10 and K39 are the binding sites which are within 5Å of the RNA molecules. I12 is among the experimentally-discovered top 20 sites, which render the highest change of fitness when mutated. D: There is no key site identified by the model when removing the structure module.

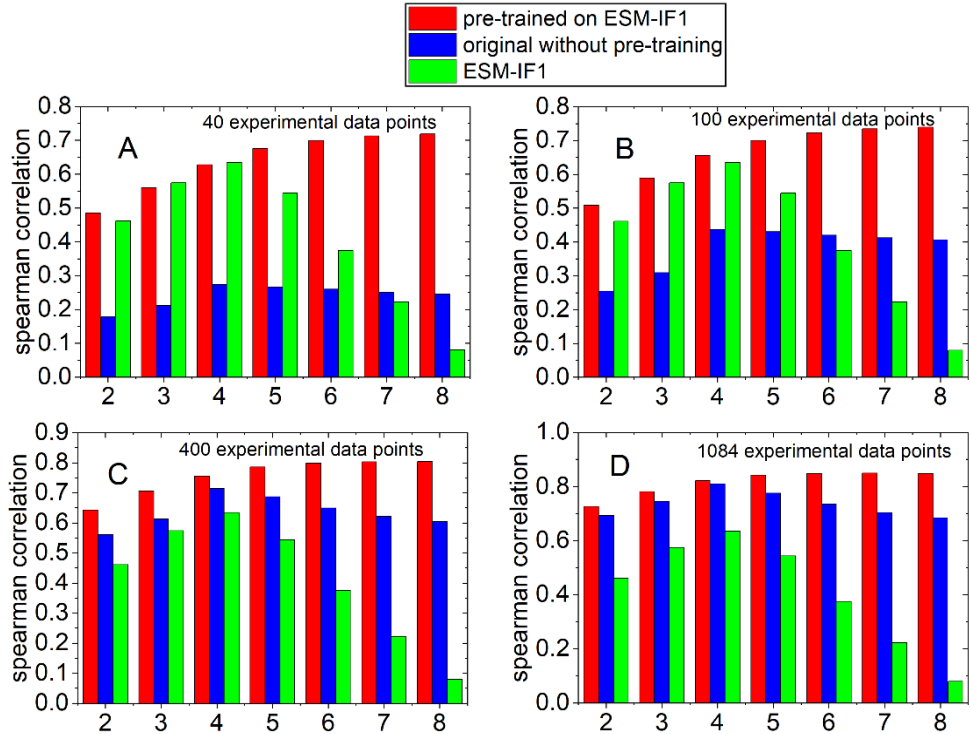

**Figure 4. Results of models trained on different number of experimental variants.** A-D: The spearman correlation of fitness prediction on multiple sites (2-8 sites) mutants by finetuning using 40, 100, 400, 1084 single-site experimental mutation results from dataset of GFP. Where the red and blue bars represent the results of the pre-trained model and the original model without pretraining, respectively. And the green bars correspond to the results of the unsupervised model ESM-IF1 as a control.

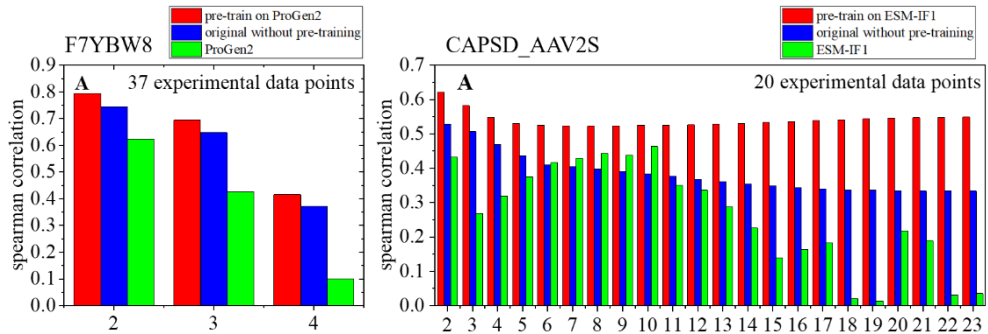

**Figure 5. Results of models trained on different datasets.** A-B: The spearman correlation of fitness prediction on high-order mutants by finetuning on 37 experimental single-site mutation results from datasets of F7YBW8 and on 20 experimental single-site mutation results of CAPSD_AAV2S, respectively. Where the red and blue bars represent the results of the pre-trained model and the original model without pretraining. And the green bars correspond to the results of the unsupervised model, which is ProGen2 for F7YBW8 and ESM-IF1 for CAPSD_AAV2S, respectively.

## Discussion

In this study, we present a supervised deep learning model, which leverages the information of both sequence and structure of protein to predict the fitness of variants. And this model is found to outperform the existing state-of-the-art ones for protein engineering. Moreover, we proposed a data augmentation strategy, which pretrains our model using the results predicted by other unsupervised model, and then finetunes the model with only a small number of experimental results. We demonstrated that such data augmentation will significantly improve the accuracy of the model when the experimental results are very limited (~40), and also for high-order mutants with >4 mutation sites. We noted that our work, especially the data-augmentation strategy proposed here, will be of great practical importance as the experimental effort it requires is generally affordable by an ordinary biochemical research group and can be applied on most protein.

## Method

### Details of Model Architecture

*Local encoder.* Residue interdependencies are crucial to evaluate if a mutation is acceptable. Several models, including ESM-MSA-1b [37], DeepSequence[14], EVE[38] and the Potts model[27], such as EVmutation[16] and ECNet [39], utilize multiple sequence alignment (MSA) to dig the constraints of evolutionary process in the residues level. In the present work, we use Potts model to establish the local encoder. This method first searches for the homologous sequences and builds MSA of the given protein with HHsuite [40]. After that, a statistical model is used to identify the evolutionary couplings by learning a generative model of the MSA of homologous sequences using a Markov random field. In the model, the probability of each sequence depends on an energy function, which is defined as the sum of single-site constraints $e_i$ and all pairwise coupling constraints $e_{ij}$:

$$E(x) = \sum_i e_i(x_i) + \sum_{i \neq j} e_{ij}(x_i, x_j) \quad (1)$$

Where $i$ and $j$ are position indices along the sequence. The $i$-th amino acid $x_i$ is encoded by a vector, in which elements are set to the single-site term $e_i(x_i)$ and pairwise coupling terms $e_{ij}(x_i, x_j)$ for $j=1,\ldots,n$, $n$ is the number of residues in the sequence. These coupling parameters $e_i$ and $e_{ij}$ can be estimated using regularized maximum pseudolikelihood algorithm [41, 42]. As the result, each amino acid in the sequence is represented by a vector whose length is $(L + 1)$, and the whole input sequence is encoded as a matrix whose size is $(L + 1) \times L$. Since the length of the local evolutionary representation of each amino acid is close to the length of the sequence, the $(L + 1)$-long vector would be transformed into a new vector with fixed length $d_l$ (in our local encoder, $d_l$=128) through a fully connected layer to avoid the overfitting issue. Sequence of protein would also pass a Bi-LSTM layer and be transformed into an $L \times d_l$ matrix for random initialization. By concatenating two matrices above, we obtain the output of local encoder $\boldsymbol{e'} =< \boldsymbol{e'_1}, \boldsymbol{e'_2}, \ldots \boldsymbol{e'_L} >$, whose size is $L \times 2d_l$.

***Global Encoder.*** Recently, the large scale pre-trained models have been successfully applied in diverse tasks for inferring protein structure or function based on sequence information. Such as prediction of secondary structure, contact prediction and prediction of mutational effects. Thus, we take a pre-trained protein language model as the global encoder which is responsible to extract biochemical properties and evolution information of the protein sequences. There are some effective language models such as UniRep [12], TAPE [43], ESM-1v [44], ESM-1b [37], ProteinBERT[11] etc. We test these language models on our validation datasets, and results show that ESM-1b performs better than others. Therefore, we chose to use ESM-1b as the global encoder. The model is a bert-based [45] context-aware language model for protein, trained on the protein sequence dataset of UniRef 50 (86 billion amino acids across 250 million protein sequences). Due to its ability to represent the biological properties and evolutionary diversity of proteins, we utilize this model as our global encoder to encode the evolutionary protein sequence. Formally, given a protein sequence $x = <x_1, x_2, ..., x_L> \in L^N$ as input, where $x_i$ is the one-hot representation of $i_{th}$ amino acids in the evolutionary sequence, $L$ is the length of the sequence, and $N$ is the size of amino acids alphabet. The global encoder first encodes each amino acid and its context to $g = <g_1, g_2, ..., g_L>$, where $g_i \in R^n$, (in ESM-1b, $n = 1420$). Then $g_i$ is projected to $g'_i$ of a hidden space $R^h$ with a lower dimension (in our default model configuration, $h = 256$), $g'_i = W_G g_i + b$, where $W_G \in R^{n \times h}$ is a learnable affine transform parameter matrix and $b \in R^h$ is the bias. The output of global encoder is $g' = <g'_1, g'_2, ... g'_L> \in R^{L \times h}$. We integrate the ESM-1b architecture into our model i.e.; we update the parameters of ESM-1b dynamically during the training process.

***Structure module.*** Structure module utilizes the microenvironmental information to guide the fitness prediction. In this part, we use the ESM-IF1 model [23] to generate the scores of mutant sequences, which evaluate their ability to be folded to the wildtype structure of the given protein. Higher scores mean these mutations are more favorable than others. Specifically, all possible single mutants at each position of a sequence would obtain the corresponding scores. The prediction sequence distribution is an $(L \times 20)$ matrix. Then we calculated the cross-entropy at each position of the sequence between the matrix above and one-hot encoding matrix of mutant sequence. After passing the results through a SoftMax function, we obtained an $(L \times 1)$ output vector, which is the reconstruction perplexities $p' = <p'_1, p'_2, ... p'_L>$ align the evolutionary sequence. In the present work, we do not directly encode distance map or the 3D coordinate of mutated protein. Since before that encoding process, we need to fold every specific mutant from their sequences, which will lead to unaffordable computational cost and is unpractical for the task of fitness prediction.

***Intra-Attention.*** The outputs of local encoder and global encoder are embedding vectors, aligning all positions of input sequence. We utilize intra-attention mechanism to compress the whole embeddings to a context vector. The inputs of attention layer are: (1) the global representations $g' = <g'_1, g'_2, ... g'_L>$ (2) the local representations

$e' =< e'_1, e'_2, ... e'_L >$ (3) the reconstruction perplexities $p' =< p'_1, p'_2, ... p'_L >$. Firstly, the local representations and global representations are normalized by layer normalization [46] over the length dimension respectively for stable training. That is, $g' = LayerNorm(g')$ and $e' = LayerNorm(e')$. Secondly, the normalized global representations and local representations are concatenated to joint-representations $r =< r_1, r_2, ... r_L >$, where $r_i = [g'_i; r'_i] \in R^{2h}$. Then we use an dot attention layer to compute the sequence attention weights $a =< a_1, a_2, ..., a_L > \in R^L$, where $a_i \in R$ is the attention weight on the $i_{th}$ position, $a_i = \frac{\exp(r_i \cdot W_a r_i)}{\sum_{k=1}^{n} \exp(r_k \cdot W_a r_k)}$, $W_a \in R^{h \times 1}$ is the learnable parameter. Besides the sequence attention weights, there is structure attention weights called structure attention $s =< s_1, s_2, ..., s_L > \in R^L$, which are calculated by reconstruction perplexities, $s_i = \frac{\exp(p'_i)}{\sum_{k=1}^{n} \exp(p'_k)}$. We use the average of sequence attention and structure attention as the final combined attention weights, that is $w =< w_1, w_2, ..., w_L >$, where $w_i = \frac{a_i + s_i}{2}$. According to the combined attention weights, we get the context vector $c = \sum_{i=1}^{L} w_i r_i$ as the embedding vector of the entire sequence.

*Output layer.* The input of output layer is the context vector $c$ from the output of attention aggregator, and an evolutionary score $d$ from the unsupervised model [23]. While the evolutionary score may not be trusted in many cases, we use a dynamic weight to take the score into account. The context vector $c$ was firstly transformed to a hidden vector $h$, where $h = ReLU(W_h c + b)$, $W_h$ and $b$ are learnable parameters, and **ReLU** [47] is the activation function. Then, the hidden vector $h$ is used to calculate the weight $p \in (0,1)$ on $d$: $p = Sigmoid(W_p[h; d])$. The scale of $p$ quantifies how much should the model trust the score from the zero-shot model. At last, we use a linear layer to compute a fitness score $y_q \in R$ according to the hidden vector $h$ directly, where $y_q = W_q h + b$. The output of our model, i.e., the prediction fitness $y \in R$ is computed as:
$$y = (1 - p) \times y_p + p \times y_q. \qquad (2)$$
We utilize the mean square error (MSE) as the loss function to update model parameters during back-propagation:
$$loss = \frac{1}{N} \sum_{i=1}^{N} (t_i - y_i)^2 \qquad (3)$$
, where $N$ is the number of samples in a mini-batch, $t_i$ is the target fitness and $y_i$ is the output fitness.

**Dataset and experimental settings**
*Benchmark dataset collection.* We first collected 20 multiple deep mutational scanning datasets from Ref [14]. Most of them only contain the fitness data of single-site mutants, while one of them (RRM)[31] also provides data of high-order mutants. The fitness data measured in these datasets include enzyme function, growth rate, peptide binding, viral replication and protein stability. We also collected the mutant data of the WW domain of human Yap1, GB1 domain of protein G in *Streptococcus sp. group G* and FOS-JUN heterodimer from Ref [48], and the prion-like domain of TDP-43 from Ref [49] to evaluate the ability of our model to predict the effect of double-sites mutant by learning from the data of single-site mutant. Besides, the ability to predict the fitness of higher order mutants (larger than 2) is tested in the dataset from Ref [29]. This study analyzed the local fitness landscape of the green fluorescent protein from *Aequorea victoria* (avGFP) by measuring the native function (fluorescence) of tens of thousands of derivative genotypes of avGFP. The detailed information on these datasets are provided in Table 4 in the Supplement Information.

***Prediction of single-site mutation effects***. We compared our model to ECNet, ESM-1b, ESM-1v and MSA transformer model on the DMS datasets. For the supervised models (ECNet and ESM-1b), we performed five-fold cross-validation on these datasets, and 12.5% of each train set are randomly selected as valid set. Spearman correlation was used to evaluate the performances of different models.

***Prediction of High-order mutation effects***. We evaluated the performance for predicting the fitness of high-order mutants by the model trained on low-order mutants. The training set for the prediction of double-site mutants only contains the experimental fitness of single-site mutants. The models used to predict the fitness of quadruple mutants of avGFP are trained on single, double, triple, and all the three types of mutants, respectively. Both in the prediction of effect of double mutants and quadruple mutants, we chose 10% of the high-order mutant data as valid set. The performances of models were evaluated by Spearman correlation.

***Data-augmentation strategy.*** The data augmentation was conducted by pre-training our model on the results predicted by the unsupervised model. To be specific, we first built a mutant library, which contains all of the single-site mutants and 30,000 double-site mutants randomly selected from tens of millions of saturated double-site mutants. Then, we used ESM-IF1 (or ProGen2) to score all of these sequences. Those sequence-score data were used to pre-train our model. While we used 90% of the data as training test, 10% as validation set. After that, we finetuned the pre-trained model on single-site mutants from experiment with the high-order mutants as test set.

**Training details.** SESNet was trained with adam optimizer with weight decay (equals to L2 norm). Hyperparameters of the model were tuned with a local grid search on the validation set. Since conducting 5-fold cross-validation and grid search on 20 datasets is costly, we only searched on two representative datasets. We performed grid search on GFP dataset for multi-sites dataset and RRM dataset for single-site dataset to obtain the best hyperparameters configuration and apply the search results in other datasets. We tested the hidden size of [128, 256, 512], learning rate of [1e-3, 5e-4, 1e-4, 5e-5, 1e-5], and dropout of [0.1, 0.2, 0.4]. Table 7 in SI shows the details of the hyperparameters configuration. All experiments are conducted on a GPU server with 10 RTX 3090 GPUs (24GB VRAM) and 2 Intel Gold 6226R CPUs with 2TB RAM.

*Model contrast*. The source code of ECNet model for contrast is downloaded from the GitHub website (https://github.com/luoyunan/ECNet) provided by Ref [17]. The ESM-1b model is also reproduced in our local computers with architecture that is described in their publication [6]. The code of ESM-IF1, ESM-1v and MSA transformer (ESM-MSA-1b) are got from the GitHub website of Facebook research (https://github.com/facebookresearch/esm). For each assay, all experiments of three different models are performed in the same dataset.

## Ethical Approval

Not applicable

## Competing interests

The authors declare no competing interests.

## Authors' contributions

LH and PT designed this project, PT and ML proposed this model, ML, LQ, YX, YGW and GF implemented the method, performed the calculations. All of the authors read and approved the final manuscript.

## Author's notes

†These authors contributed equally to this work.

*To whom correspondence should be addressed: tpan1039@alumni.sjtu.edu.cn; hongl3liang@sjtu.edu.cn.

## Funding

This work was financially supported by the Natural Science Foundation of China (Grant No. 12104295, 11974239, 31630002, 61872094, 61832019), the Innovation Program of Shanghai Municipal Education Commission, and Shanghai Jiao Tong university Multidisciplinary research fund of medicine and engineering YG 2016QN13. The computing hardware resource was supported by the Center for High Performance Computing at Shanghai Jiao Tong University.

## Availability of data and materials

Source code for SESNet and all the datasets used in the present work can be found in the supplemental materials. Where the original sources of datasets have been declaimed and cited in the main text.